\documentclass{aa}

\usepackage{psfig,graphics}
\begin{document}

   \thesaurus{11  
              (11.01.2;   
               11.17.2;   
               11.17.3;   
               11.17.4)}  
               
   \title{Two new quasars at $z$ = 1.90 and $z$ = 0.15 from the 
          Cal\'an--Tololo Survey
 \thanks{Based on observations collected at the European Southern 
        Observatory, La Silla, Chile}}


   \author{J.~Rossa
         \inst{1}
         \and
         C.~Tappert
          \inst{1}
          \and   
         Th.~Augusteijn
          \inst{2}
          \and          
         J.~Maza
          \inst{3}}
   \offprints{jrossa@astro.ruhr-uni-bochum.de}

   \institute{Astronomisches Institut der Ruhr-Universit\"at Bochum,
             D--44780 Bochum, Germany
         \and
             European Southern Observatory, Alonso de C\'ordova 3107, 
             Vitacura, Santiago, Chile
         \and
             Departamento de Astronom\'{\i}a, Universidad de Chile, 
             Casilla 36--D, Santiago, Chile}
             

   \date{Received <date> / Accepted <date>}

   \maketitle

   \begin{abstract}

We present the discovery of two new quasars, that initially appeared as 
candidate cataclysmic variables (CVs) in a list of the Cal\'an--Tololo 
objective prism survey. However, additional spectroscopic and photometric 
observations revealed their true nature. Furthermore we present data for a 
third quasar, CTCV J0322--4653, which has also been found in our study, but 
which has already been included in the list of Maza et al.~(\cite{Ma96}) as 
CTS\,0639. Here we present the spectra for the first time.

   \end{abstract}

%

\section{Introduction}

Although large surveys aiming at the detection of bright QSOs have been 
undertaken such as the photographic Palomar-Green survey (Green et al. \cite{
Gr86}), which was carried out in the early seventies, and the more recent 
Hamburg/ESO survey (e.g.,\,Wisotzki et al. \cite{Wi}) and the Cal\'an--Tololo 
survey (e.g.,\,Maza et al. \cite{Ma96}), serendipitous discoveries are being 
reported from time to time. The recent discovery of a bright and nearby QSO 
less than $1^\circ$ away from 3C273, one of the best studied quasars (Read 
et al. \cite{ReMi}) is such a case. 

Here we present the spectra of three new quasars from the Cal\'an--Tololo 
Survey. Somewhat ironically, these objects were initially selected as 
candidate cataclysmic variables on the basis of their visual appearance on the 
objective prism plates (Augusteijn et al. \cite{Au99}). Their blue fluxes 
and a strong emission line near the H$\beta$ rest wavelength led to such 
misidentifications. The follow--up observations, however, revealed their true 
nature. 


\section{Observations and Data Reduction}

The spectroscopic and photometric observations have been carried out with 
various telescopes at ESO, La Silla, Chile (see Tab.~\ref{T1}). The data 
reduction was performed in the usual manner including bias level subtraction 
and flatfielding using the various IRAF packages.

\begin{table*}
 \caption[]{Journal of observations}
 \label{T1}
 \begin{flushleft}
  \begin{tabular}{lllllll}
  \noalign{\smallskip}
  \hline
  \hline
  Object designation & Date & Telescope & Instrument & $\Delta\lambda$\,
[{\AA}] & $\Delta\lambda_{\rm{FWHM}}$\,[{\AA}] & $t_{\sf int}$\,[sec] \\
  \hline
  \noalign{\smallskip}
  CTCV J0322--4653 & 31/07/96 & ESO/MPI 2.2m & EFOSC2 & 3250--10250 & 30 & 
1800 \\
   & 02/10/96 & Dutch 0.9m & Direct & B Bessel & -- &2$\times$200 \\
   & & Dutch 0.9m & & V Bessel & -- & 113$\times$90 \\
   & & Dutch 0.9m & & R Bessel & -- & 2$\times$90 \\
  CTCV J1322--2101 & 12/04/96 & Danish 1.5m & DFOSC & 5250--9250 & 13 & 600 \\
   & 25/04/98 & ESO/MPI 2.2m & IRAC2b & $\rm{K'}$ & -- & 600 \\
  CTCV J1329--1920 & 03/08/96 & ESO/MPI 2.2m & EFOSC2 & 3500--10000 & 30 & 
3600 \\
  \noalign{\smallskip}
  \hline
 \end{tabular}
 \end{flushleft}
\end{table*}


\section{Results}

\begin{table*}
  \caption[]{Positions, photometrically and spectrophotometrically derived 
magnitudes$^1$, color--indices, and derived redshifts}
  \label{T2} 
  \begin{flushleft}
 \begin{tabular}{llllllll}
  \noalign{\smallskip}
  \hline
  \hline
  Object name & R.A. (J2000) & Dec. (J2000) & Magnitude & 
  $B-V$ & $V-R$ & Method & $z$\\
  \hline
  \noalign{\smallskip}
  CTCV J0322--4653 & 03$^{\rm h}$22$^{\rm m}$28\fs53 & $-$46\degr53\arcmin
  01\farcs8 & $m_V =$ 17.89$\pm$0.02 & 0.37$\pm$0.01 &
  0.31$\pm$0.01 & phot & 1.21\\
   & & & $m_V =$ 17.84$\pm$0.06 & 0.37$\pm$0.01 & 0.29$\pm$0.01 & spec \\
  CTCV J1322--2101 & 13$^{\rm h}$22$^{\rm m}$18\fs03 & $-$21\degr01\arcmin
  41\farcs9 & $m_R =$ 16.39$\pm$0.06 & -- & -- & spec & 0.15\\
  CTCV J1329--1920 & 13$^{\rm h}$29$^{\rm m}$16\fs91 & $-$19\degr20\arcmin
  02\farcs8 & $m_V =$ 18.78$\pm$0.06 & 0.48$\pm$0.01 & 0.24$\pm$0.01 & 
  spec & 1.90\\
  \noalign{\smallskip}
  \hline
 \multicolumn{8}{l}{\footnotesize 1) Spectrophotometric magnitudes have been
 corrected for an average slit loss of 0.65\,mag}\\[-0.2cm]
 \end{tabular}
 \end{flushleft}
 \end{table*}

\begin{figure}[h]
\rotatebox{270}{\resizebox{5.8cm}{!}{\includegraphics{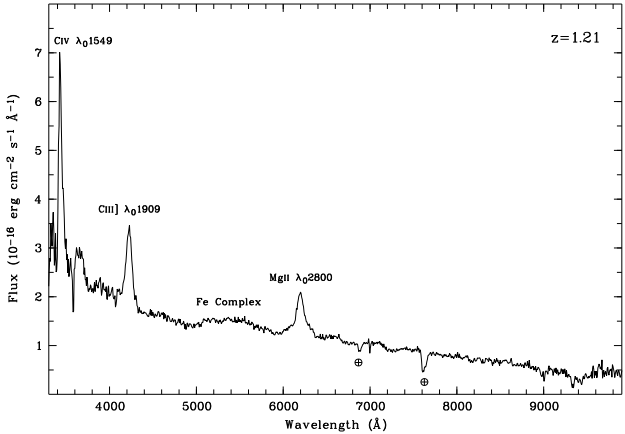}}}
\caption[]{Spectrum of CTCV J0322-4653. Telluric features are marked 
by $\oplus$.}
\label{F2a}
\end{figure}

\begin{figure}[h]
\rotatebox{270}{\resizebox{5.8cm}{!}{\includegraphics{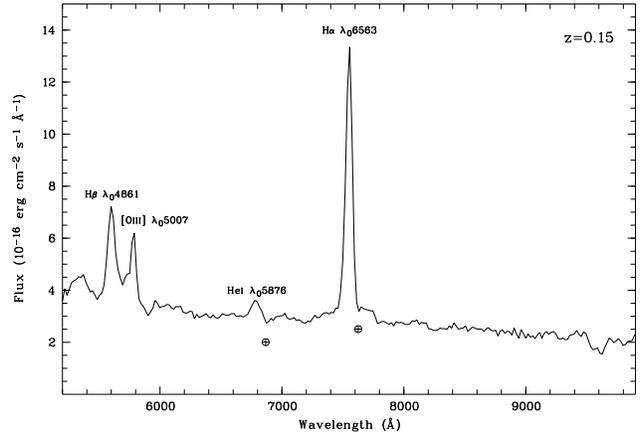}}}
\caption[]{Spectrum of CTCV J1322-2101. Telluric features are marked 
by $\oplus$.}
\label{F2b}
\end{figure}

\begin{figure}[h]
\rotatebox{270}{\resizebox{5.8cm}{!}{\includegraphics{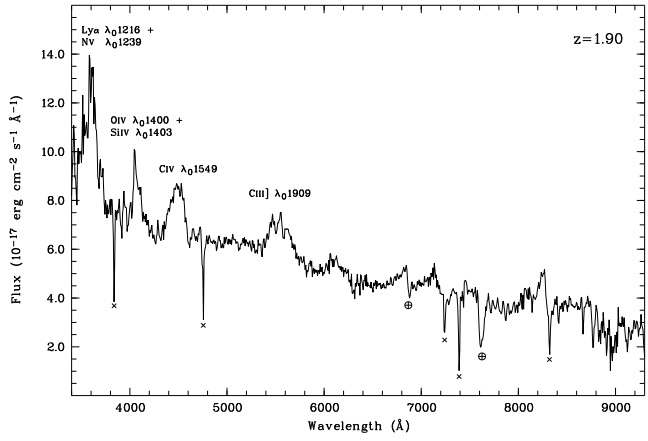}}}
\caption[]{Spectrum of CTCV J\,1329-1920. Telluric features are marked 
by $\oplus$, additional artifacts are marked by $\times$.}
\label{F2c}
\end{figure}

The inspection of our spectra revealed broad emission lines typical for 
non--stellar objects (Fig.~\ref{F2a}--\ref{F2c}). However, an analysis of the 
FWHMs, in both the DSS frames and our photometric data, yielded point source 
characteristics for all three targets, letting us suspect a QSO nature.
Also the measured color indices (Tab.~\ref{T2}) show values typical for 
quasars. The redshifts have been measured from several emission lines 
after identifying one reference emission line in each spectrum. The remaining 
lines were identified afterwards based on the comparison with the known rest 
wavelengths for the one low-- and the two intermediate--redshifted objects 
(see Tab.~\ref{T2}). 

For the low redshift QSO, CTCV J1322--2101, only an R magnitude could be 
determined due to the limited spectral range covered. However, assuming an 
upper color limit of $V-R$ $\approx$ 0.4, this indicates that the object 
is probably brighter than $M_V = -$23.

Motivated by the low redshift value of CTCV J 1322--2101, additional 
near--infrared data were obtained in the K$'$ band in order to check whether 
the host galaxy could be detected. No evidence for any extended emission 
around the object at the level of sensitivity was found. Furthermore, the 
measured FWHM from the photometry was consistent with stellar values and much 
smaller than the FWHM of the faint galaxy that is located $\approx 14''$ SE 
of the QSO.

From all studied quasars discovered by the Cal\'an--Tololo objective prism 
survey, only very few quasars with redshifts $z \leq$ 0.3 have been found. 
Our finding would be the only high--luminosity quasar ($M_B < -23$) with 
$z \leq$ 0.2 in this survey (Maza et al. \cite{Ma96}, and references therein). 


\vspace{0.0cm}
\vspace{0.0cm}

\vspace{0.0cm}
\vspace{0.0cm}

\vspace{0.0cm}
\vspace{0.0cm}

\begin{acknowledgements}
Some of the data were obtained, and reduced during a research stay of CT at 
the Universidad Cat\'olica, Santiago, Chile. This was financially supported 
by the Deutscher Akademischer Austauschdienst (DAAD) under grant D/94/14720.
We would also like to thank the referee, Dr. Lutz Wisotzki, for helpful 
comments.
\end{acknowledgements}


\end{document}